\documentclass[a4paper]{jpconf}
\usepackage{graphicx}
\usepackage{bm}

\begin{document}

\title{Implications of nonzero $\theta_{13}$ for the neutrino mass hierarchy$^*$}

\author{\underline{D. J. Ernst}$^{1,2}$, B. K. Cogswell$^{1,2}$, H. R. Burroughs$^1$, J. Escamilla-Roa$^1$, D. C. Latimer$^3$}

\address{$^1$ Department of Physics and Astronomy, Vanderbilt University, Nashville, Tennessee 37235}
\address{$^2$ Department of Physics, Fisk University, Nashville, Tennessee 37208}
\address{$^2$ Department of Physics, University of Puget Sound, Tacoma, Washington, 98416}

\ead{david.j.ernst@vanderbilt.edu}

\begin{abstract}
The Daya Bay, RENO, and Double Chooz experiments have discovered a large non-zero value for $\theta_{13}$. We present a global analysis that includes these three experiments, Chooz, the Super-K atmospheric data, and the $\nu_\mu \rightarrow \nu_e$ T2K and MINOS experiments that are sensitive to the hierarchy and the sign of $\theta_{13}$. We report preliminary results in which we fix the mixing parameters other than $\theta_{13}$ to those from a recent global analysis. Given there is no evidence for a non-zero CP violation, we assume $\delta=0$. T2K and MINOS lie in a region of $L/E$ where there is a hierarchy degeneracy  in the limit of $\theta_{13}\rightarrow 0$ and no matter interaction. For non-zero $\theta_{13}$, the symmetry is partially broken, but a degeneracy under the simultaneous exchange of both hierarchy and the sign of $\theta_{13}$ remains. Matter effects break this symmetry such that the positions of the peaks in the oscillation probabilities maintain the two-fold symmetry, while the magnitude of the oscillations is sensitive to the hierarchy. This renders T2K and NO$\nu$A, with different baselines and different matter effects, better able in combination to distinguish the hierarchy and the sign of $\theta_{13}$. The present T2K and MINOS data do not distinguish between hierarchies or the sign of $\theta_{13}$, but the large value of $\theta_{13}$ yields effects from atmospheric data that do. We find for normal hierarchy, positive $\theta_{13}$, $\sin^22\theta_{13}=0.090\pm0.020$ and is 0.2\% probable it is the correct combination; for normal hierarchy, negative  $\theta_{13}$, $\sin^22\theta_{13}=0.108\pm0.023$ and is 2.2\% probable; for inverse hierarchy, positive  $\theta_{13}$, $\sin^22\theta_{13}=0.110\pm0.022$ and is 7.1\% probable; for inverse hierarchy, negative  $\theta_{13}$, $\sin^22\theta_{13}=0.113\pm0.022$ and is 90.5\% probable, results that are inconsistent with two similar analyses.
\end{abstract}

\section{Introduction}
Neutrinos undergo the phenomenon of flavor oscillations; a neutrino created in a particular flavor (electron, mu, or
tau) will change its flavor, a phenomenon not accounted for in the standard model of the electroweak interaction. A
possible explanation of this behavior is neutrino oscillations. The neutrinos are assigned masses. The created
neutrino has definite flavor, but these flavor eigenstates are not the physical free particles, the mass
states. This leads to a phenomenology that is parameterized by three mixing angles, a CP violating phase, and two mass-squared differences, $\Delta_{ij}=:m_i^2-m_j^2$.
\vspace{6pt}

\noindent $^*$to appear in Horizons of Innovative Theories, Experiments, and Supercomputing in Nuclear Physics (New Orleans, June 4-6, 2012)
\vfill\eject
Experiments have found that neutrino oscillations approximately break into two $2\times 2$
oscillations, termed solar and atmospheric oscillations. The solar oscillations are predominantly governed by the solar data
and the KamLAND experiment, and they predominantly determine $\theta_{12}$ and $\Delta_{21}$, while the atmospheric
oscillation parameters $\theta _{23}$ and $\Delta_{32}$ are predominantly determined by the atmospheric data and the long-baseline muon disappearance data. The last mixing angle $\theta_{13}$ and the CP-violating phase $\delta$ mix the two $2\times 2$ oscillations. 

An outstanding question in neutrino phenomenology is that of the hierarchy, whether the small mass-squared difference lies below (normal hierarchy) or above (inverse hierarchy) the large mass-squared difference. A related question, as we will show, is whether the value of $\theta_{13}$ is positive or negative. We utilize the convention \cite{1} that $-\frac{\pi}{2}\le\theta_{13}< +\frac{\pi}{2}$ and $0\le\delta<\pi$. This is the more convenient parameterization for discussing the symmetries of interest here. Recent long-baseline reactor experiments \cite{2,3,4} have accurately measured the value of $\theta_{13}$ and found it to be large, $\sin^2(2\,\theta_{13})\approx 0.1$. Before these measurements, the world's data demonstrated very little preference  for either the hierarchy or the sign of $\theta_{13}$ \cite{5}. The goal of this work is to examine the effect that the recent knowledge of $\theta_{13}$ has on the hierarchy question and on the sign of $\theta_{13}$. This report is preliminary in that we will fix the mixing parameters except $\theta_{13}$ at values from a recent analysis. Since there is no strong evidence for CP-violation, we set the CP phase $\delta$ to zero. 
 
\begin{figure}
\includegraphics*[width=3in]{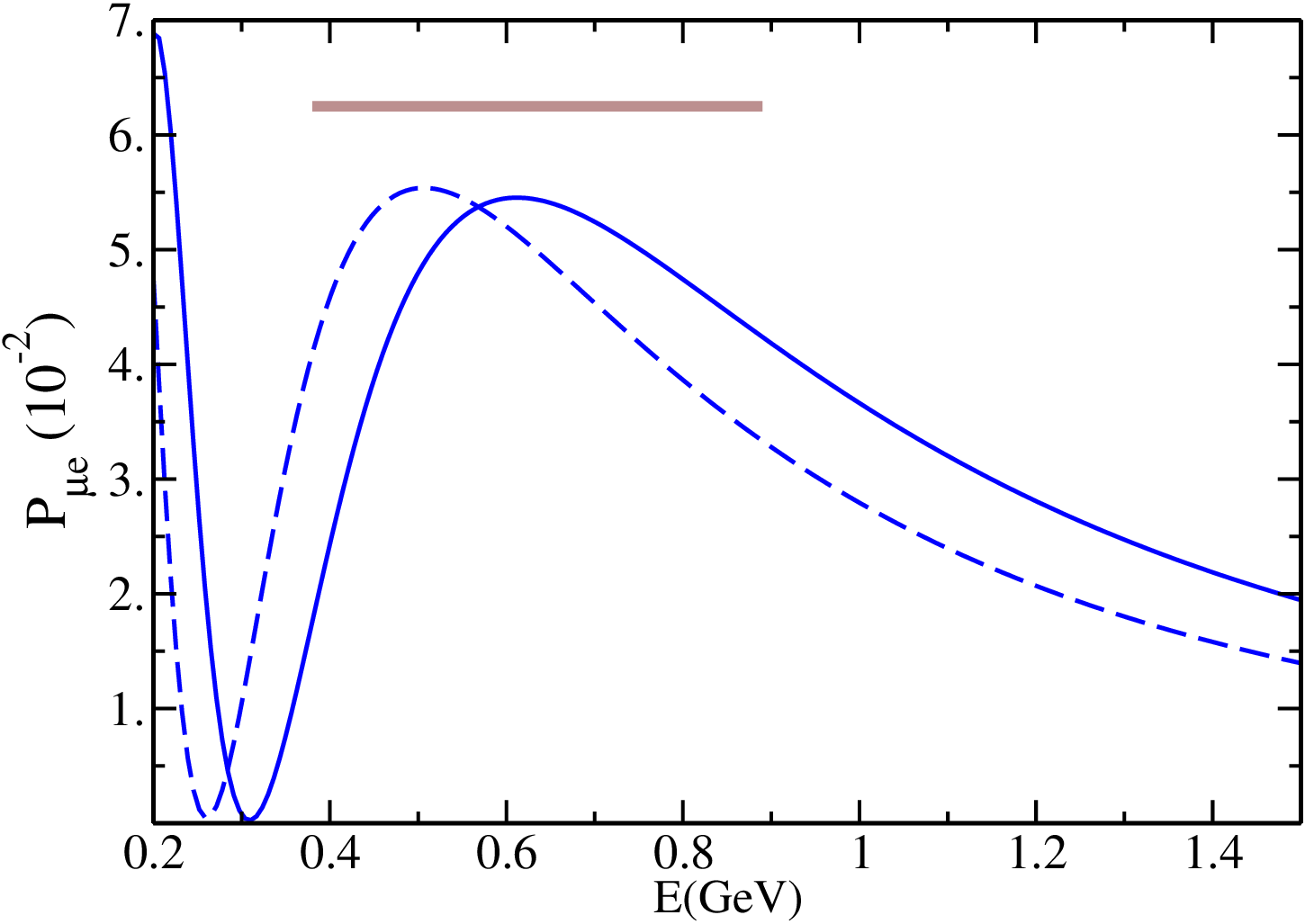}
\hfill\begin{minipage}[b]{2.9in}\caption
{${\mathcal P}_{\mu e}$  in vacuum versus neutrino energy for the T2K experiment. The solid curve depicts the normal hierarchy, positive $\theta_{13}$ as well as the inverse hierarchy, negative $\theta_{13}$ results. The dashed curve depicts the normal hierarchy, negative $\theta_{13}$ as well as  the inverse hierarchy, positive $\theta_{13}$ results. The horizontal bar represents the range where the muon interaction rate is greater than 50\% of its maximum. }
\end{minipage}
\label{fig1}
\end{figure}

\section{Symmetries and Degeneracies}
The task of extracting neutrino mixing parameters from data is complicated by the existence of symmetries that lead to degeneracies \cite{6}, where degeneracies mean different mixing parameters yield the same oscillation probability. The MINOS \cite{7}, T2K \cite{8}, and the future NO$\nu$A \cite{9} experiments are at an $L/E$ where such a symmetry nearly exists. In the limit of $\theta_{13}=0$ and no matter effects, these experiments would be insensitive to the hierarchy. The symmetry pertinent to these experiments, using our bounds on the mixing angles, is the combined  interchange of hierarchy and a change of the sign of $\theta_{13}$, a four-fold degeneracy. In Fig.~\ref{fig1} we depict the vacuum oscillation probability ${\mathcal P}_{\mu e}$ for the T2K experiment for full three-neutrino mixing using the parameters  \cite{10} $\Delta_{21}=7.54\times 10^{-5}$ eV$^2$, $\theta_{12}=0.557$, $\Delta_{31}=2.42\times 10^{-3}$ eV$^2$, $\theta_{13}=0.1$, and $\theta_{23} = 0.674$.
\begin{figure}[t]
\begin{center}
\includegraphics*[width=3in]{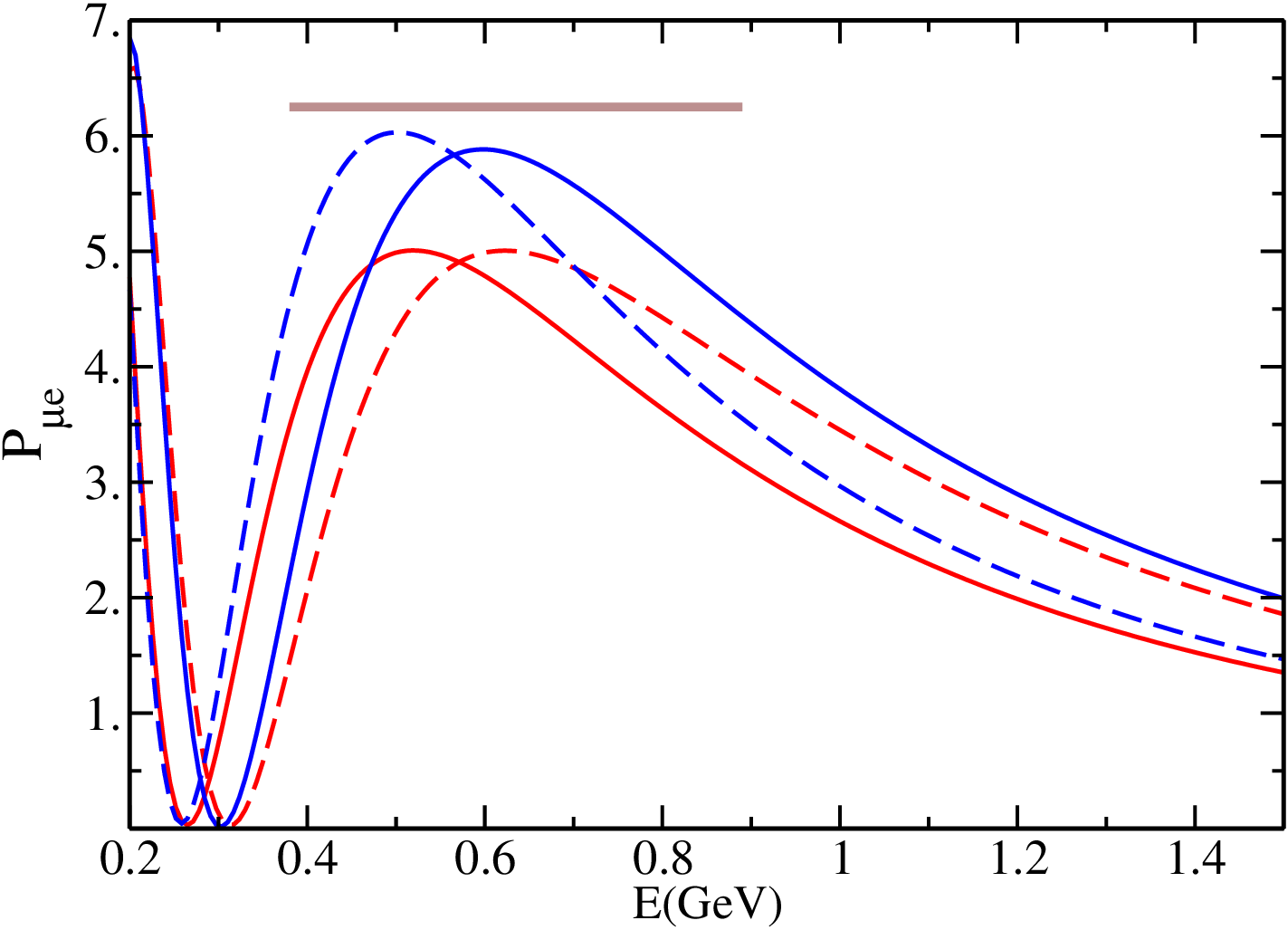}\hspace{0.1in}\includegraphics*[width=3in]{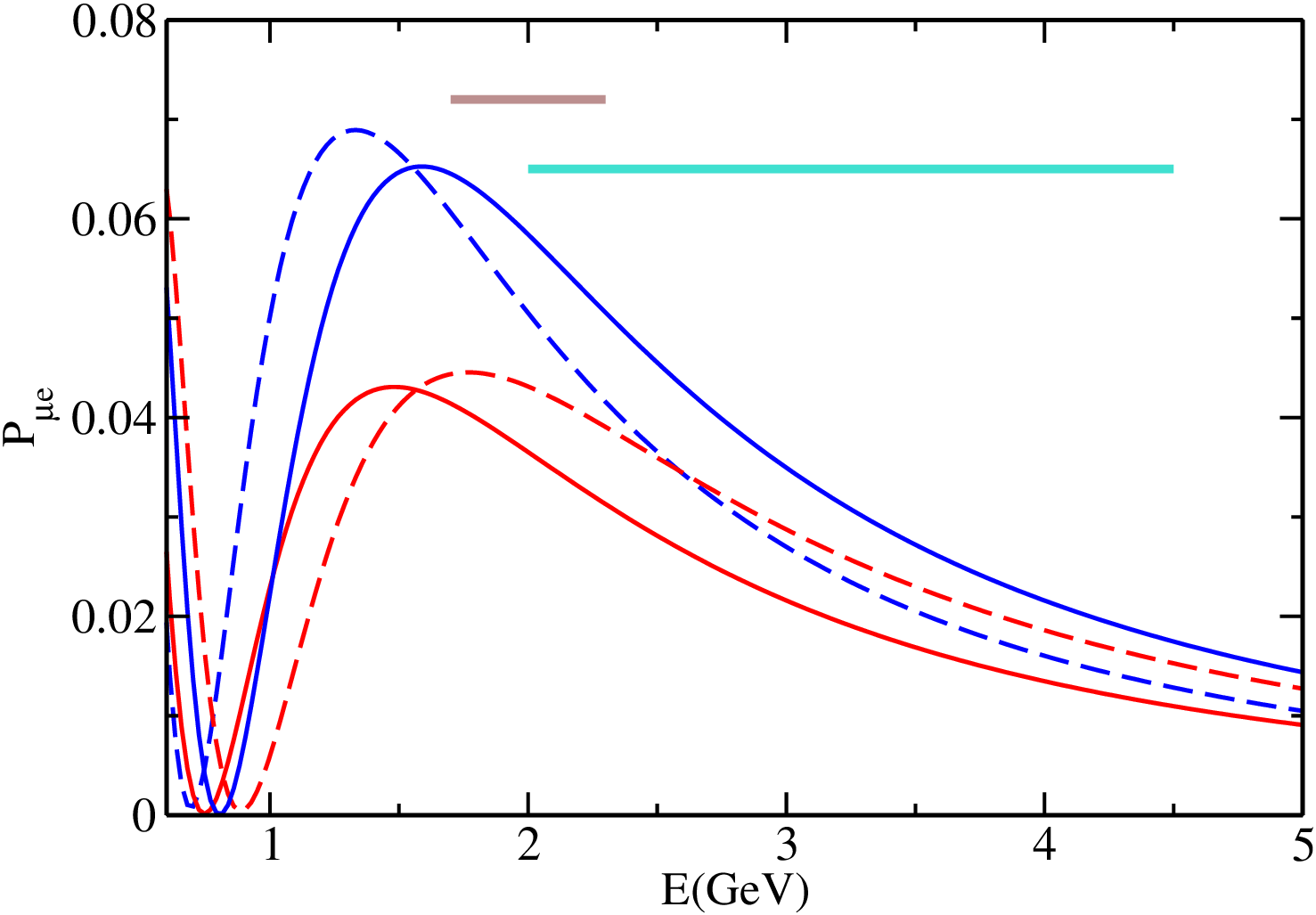}
\caption{${\mathcal P}_{\mu e}$  versus neutrino energy for the T2K experiment, left, and the MINOS/NO$\nu$A experiments, right. The blue curves depict the normal hierarchy, the red curves the inverse hierarchy. The solid curves depict positive $\theta_{13}$, the dashed curves negative $\theta_{13}$. For the left curve, the horizontal brown bar represents the range where the muon interaction rate is greater than 50\% of its maximum, while for the right curve the brown bar represents the same quantity for the NO$\nu$A experiment, and the turquoise bar represents the MINOS experiment. }
\label{fig2}
\end{center}
\end{figure}

\begin{figure}[t]
\begin{center}
\includegraphics*[width=3in]{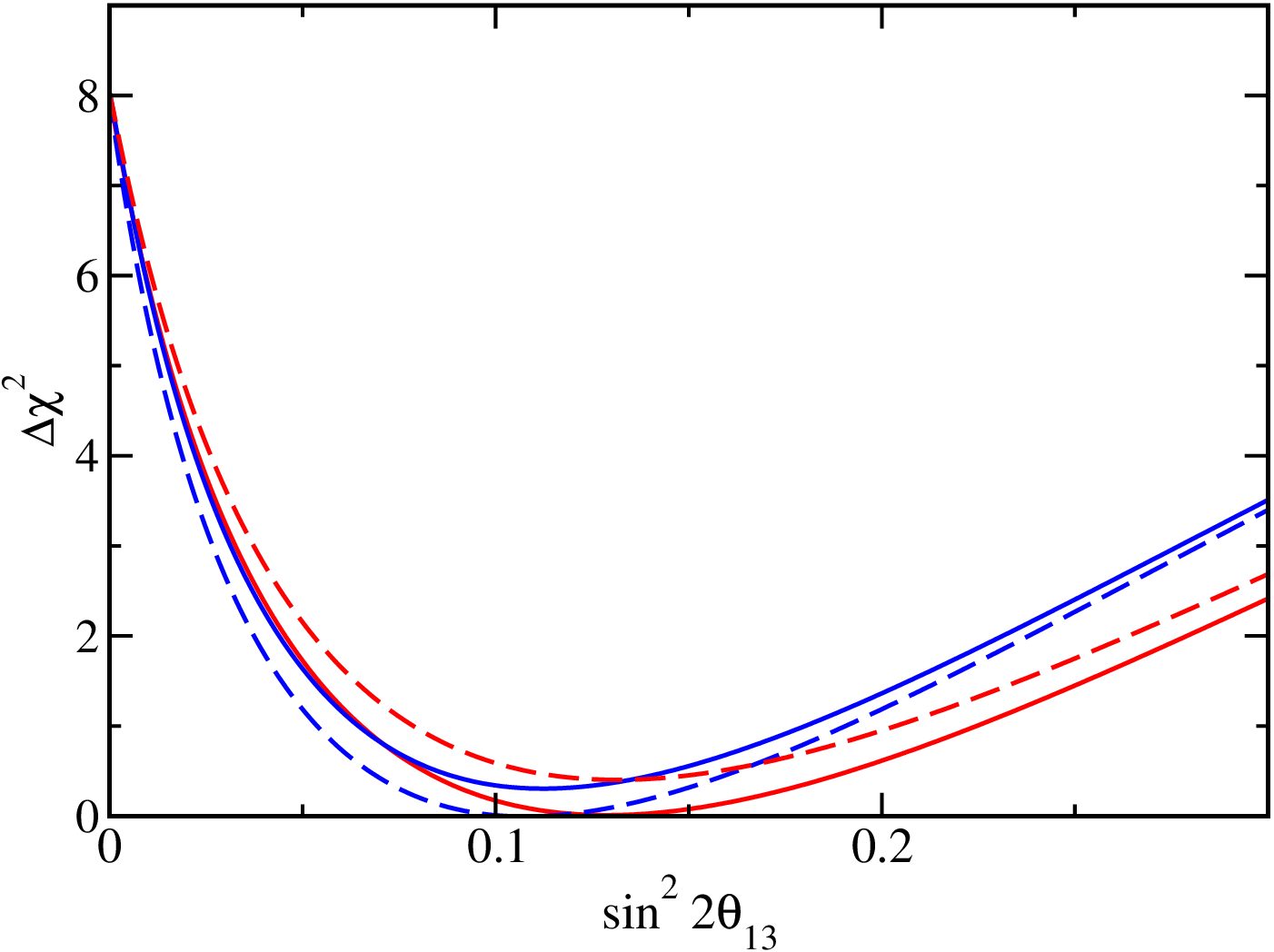}\hspace{0.1in}\includegraphics*[width=3in]{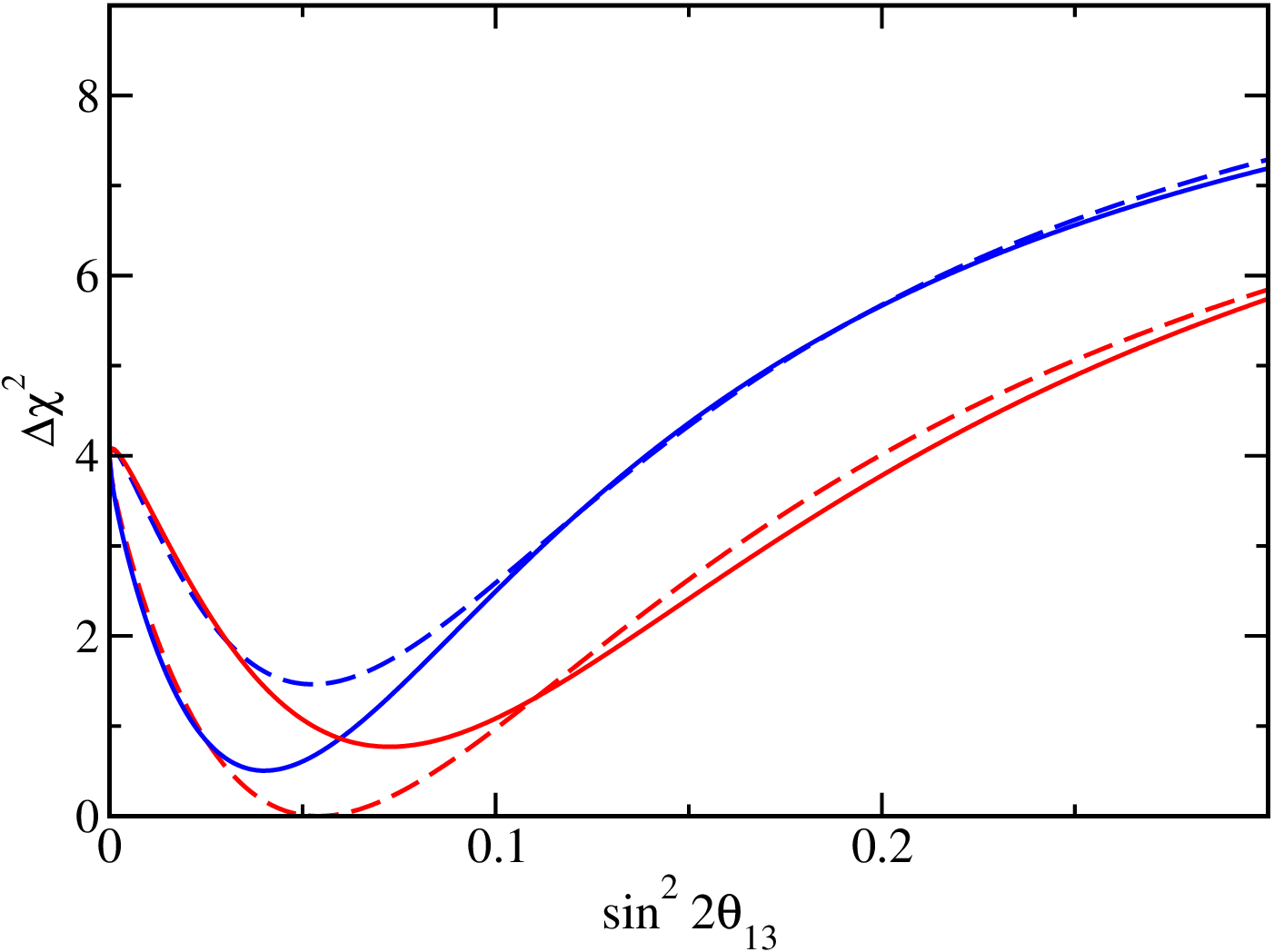}
\caption{$\Delta\chi^2$ versus $\sin^2\,2\,\theta_{13}$ for T2K, left, and MINOS, right. Solid curves are normal hierarchy, dashed inverse hierarchy. Blue curves are positive $\theta_{13}$, red curves negative $\theta_{13}$.}
\label{fig3}
\end{center}
\end{figure}

We see that the four-fold degeneracy is broken by non-zero $\theta_{13}$. However, the solid curve represents both the normal hierarchy, positive $\theta_{13}$ as well as the inverted hierarchy, negative $\theta_{13}$ results while the dashed curve represents the inverted hierarchy, positive $\theta_{13}$ and the normal hierarchy, negative          
$\theta_{13}$ results. Thus non-zero $\theta_{13}$ breaks the four fold degeneracy down to two 
\vfill\eject
\noindent two-fold degeneracies. Hierarchy and linear in $\theta_{13}$ effects require interference between the two mass-squared difference terms, and this is the smallest value of $L/E$ where such effects appear. These effects are greatest at the peak of the first oscillation generated by the small solar mass-squared differences. For vacuum oscillations this occurs \cite{11}  when $L/E = 1.6\times10^4$ km/GeV. Including matter effects yields $L=4.0\times10^3$ km and $E=0.23$ GeV as typical ideal values.  Extrapolating from the results in Ref.~\cite{12} to the now known value of $\theta_{13}$  gives a large ${\mathcal P}_{\mu e}$ with a 20\% difference between ${\mathcal P}_{\mu e}$ for $\delta=0$ and $\delta=\pi/2$ for such an experiment.   
\vfill\eject 
We present in Fig.~\ref{fig2} the ${\mathcal P}_{\mu e}$ verses neutrino energy including the MSW matter effects for T2K \cite{8}  and  MINOS \cite{7}/ NO$\nu$A \cite{9}. The effect of including the matter interaction of the neutrino is to alter the magnitude of the oscillations while leaving the position of the peaks nearly unchanged. Thus spectral information that measures the position of the peaks maintains approximately the two-fold degeneracy of the vacuum oscillations. The magnitude of the signal then breaks this symmetry to distinguish the hierarchy while it alone does not distinguish well the sign of $\theta_{13}$.

There is a complementarity between the T2K and NO$\nu$A experiments. T2K is situated to determine the peak of the oscillation probability, ideal for breaking the four-fold degeneracy. NO$\nu$A has the longer baseline and the larger matter effects thus increasing the difference in the magnitude of the oscillations, giving it the greater sensitivity to the hierarchy. 

\begin{table}[t]
\caption{Results for T2K, MINOS, and the combination. For the four cases of hierarchy and the sign of $\theta_{13}$, we present the best fit values of $\sin^22\,\theta_{13}$, its 90\% error bars, the maximum confidence level at which $\theta_{13}\ne0$, and the probability that each case is the correct choice.}
\begin{center}
\begin{tabular}{lcccccc}
\br
experiment&hierarchy&sign $\theta_{13}$&$\sin^22\,\theta_{13}$&CL, $\theta_{13}\ne0$&\% probable\\
\mr
&normal&$+$&$0.11_{-0.11}^{+0.20}$&99.1\%&22.1\%\\
T2K&normal&$-$&$0.11_{-0.09}^{+0.29}$&99.2\%&24.9\%\\
&inverse&$+$&$0.13_{-0.10}^{+0.11}$&99.4\%&28.8\%\\
&inverse&$-$&$0.13_{-0.10}^{+0.12}$&99.3\%&24.3\%\\
\mr
&normal&$+$&$0.040^{+0.190}_{-0.040}$&80.8\%&21.5\%\\
MINOS&normal&$-$&$0.053^{+0.215}_{-0.053}$&75.6\%&16.3\%\\
&inverse&$+$&$0.073^{+0.208}_{-0.073}$&83.1\%&28.0\%\\
&inverse&$-$&$0.054^{+0.194}_{-0.054}$&85.8\%&34.3\%\\
\mr
&normal&$+$&$0.064^{+0.103}_{-0.045}$&99.5\%&18.4\%\\
T2K+MINOS&normal&$-$&$0.072^{+0.064}_{0.045}$&99.5\%&17.9\%\\
&inverse&$+$&$0.093^{+0.114}_{-0.059}$&99.7\%&33.2\%\\
&inverse&$-$&$0.080^{+0.118}_{-0.055}$&99.7\%&30.5\%\\
\br
\label{tab1}
\end{tabular}
\end{center}
\end{table}

\section{Long-Baseline Experiments}
Data from T2K and MINOS $\nu_\mu\rightarrow \nu_e$ appearance experiments are available, but both lack significant statistical quality. The $\Delta\chi^2$ versus $\theta_{13}$ results are presented in Fig.~\ref{fig3}. For T2K, the left graph, we see that the quality of the fit (the value of $\Delta\chi^2$ at the minimum) is governed by the spectral information, while the position of the minimum is determined by the height of the oscillation probability and is thus correlated with the hierarchy. For MINOS, we see that the longer baseline increases the matter effects. For small $\theta_{13}$ the two fold symmetry is present. However, for the now known value of $\sin^2\,2\theta_{13} \approx 0.1$, MINOS is sensitive only to the hierarchy.  From Fig.~\ref{fig2} we see that this is because MINOS is in a region where the magnitude of the data is important but spectral information is limited as its data does not extend down in energy sufficiently far to be sensitive to the position of the peaks.
\vfill\eject
In Table~\ref{tab1} we present results from our analysis of the T2K, MINOS and combined experiments. Our analyses agree quite well with that of the experimentalists. For T2K we utilize a new estimate of the earth's matter density \cite{13} of 2.6 gm/cm$^3$ rather than the 3.2 gm/cm$^3$ used by the experimentalists. For the present T2K data we see relatively little distinction between the hierarchy or the sign of $\theta_{13}$. There is a statistically insignificant favor (disfavor) for the inverse hierarchy and positive $\theta_{13}$ (normal hierarchy and positive $\theta_{13}$). 

We see in Fig.~\ref{fig3} that the $\Delta\chi^2$ curves are not quadratics. Hence the use of normal statistics is not valid. We calculate the error bars and the probability that $\theta_{13}\ne 0$ using the Bayesian method from \cite{14} with the prior taken as a constant for the independent variable $\sin^22\theta_{13}$. Both experiments provide an indication that $\theta_{13}$ is not zero, T2K at the 99\% CL and MINOS at the 80\% CL, the combined at the 99.5\% CL. They both show a small dependence on hierarchy and the sign of $\theta_{13}$, with the combined result preferring the inverse hierarchy at 64\% probability. 

\section{Reactor Experiments}
Recent reactor experiments, Double Chooz \cite{2}, Daya Bay {\cite 3}, and RENO \cite{4}, have measured the value of $\theta_{13}$. The obtained value of $\theta_{13}$ is surprisingly  large. Insufficient information is given to perform an {\it ab initio} analysis of the Daya Bay and RENO experiments. However, we find that if we take the data to be given by the points in Fig.~4 for Daya Bay and Fig. ~3 for RENO, utilize the systematic errors given, and take the statistical errors as implied by the number of events in each detector, we  get results nearly identical to those given by the experimentalists. 

\begin{figure}[t]
\includegraphics*[width=3in]{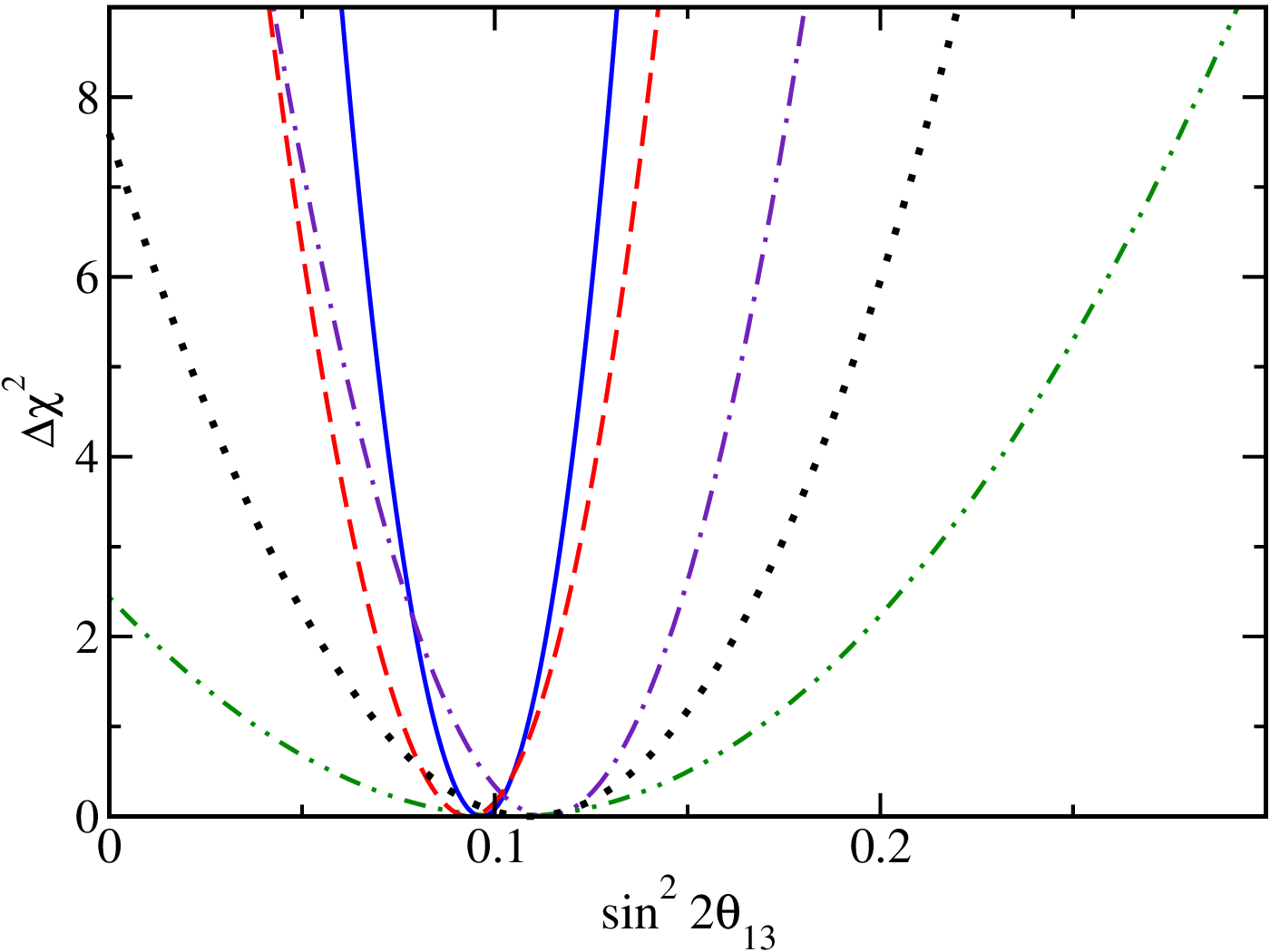}
\hfill
\begin{minipage}[b]{2.9in}
\caption
{$\Delta\chi^2$ versus $\sin^22\,\theta_{13}$ for the Chooz experiment (dot-dot-dash green curve)  with modified flux, the Double Chooz experiment (dotted black curve), the RENO experiment (indigo dot-dash), the Daya Bay experiment (red dashed), and the combination of all four experiments (solid blue).}
\end{minipage}
\label{fig4}
\end{figure}
\begin{table}[b]
\caption{Implications of the reactor experiments Chooz, Double Chooz, Daya Bay, RENO and the combination. We present the best fit values of $\sin^22\,\theta_{13}$ and its 90\% error bars, the maximum confidence level at which $\theta_{13}=0$ is excluded.}
\begin{center}
\begin{tabular}{lcccccc}
\br
experiment&$\sin^22\,\theta_{13}$&CL, $\theta_{13}\ne0$\\
\mr
Chooz&$0.104^{+0.091}_{-0.094}$&94\%&&\\
Double Chooz&$0.109^{+0.061}_{-0.064}$&99.7\%\\
Daya Bay&$0.093^{+0.026}_{-0.027}$&$1.-6.\times10^{-6}$ \%&&\\
RENO&$0.113^{+0.037}_{-0.038}$&$1.-1.\times10^{-4}$ \%\\
Combined&$0.097^{+0.019}_{-0.020}$&$1.-1.\times10^{-9}$ \%\\
\br
\label{tab2}
\end{tabular}
\end{center}
\end{table}

In Fig.~\ref{fig4} we present $\Delta\chi^2$ for these reactor experiments, plus results for the original Chooz \cite{15} experiment, in which we use an increase in the  flux by 3.3\% \cite{16}. This change makes Chooz consistent with the recent experiments. Daya Bay and RENO, by virtue of having identical near and far detectors, provide a definitive non-zero value for $\theta_{13}$. Reactor experiments measure ${\mathcal P}_{ee}$ and have no sensitivity to hierarchy or the sign of $\theta_{13}$. All four of these experiments are consistent with each other, all indicating a value for $\sin^22\,\theta_{13}\approx 0.1$.

In Table~\ref{tab2} we present results for the reactor experiments. The result for the combination is
$\sin^2 2 \theta_{13} = 0.097^{+0.019}_{-0.020}$. $\theta_{13}$ is non-zero at the confidence level of CL$=1.0-1.\times 10^{-9}$ \%, or 6.2 $\sigma$. We find a somewhat smaller significance because we use the Bayesian approach from Ref.~\cite{14}.

\section{Super-K Atmospheric Experiment}
\begin{figure}[t]
\begin{center}
\includegraphics*[width=3in]{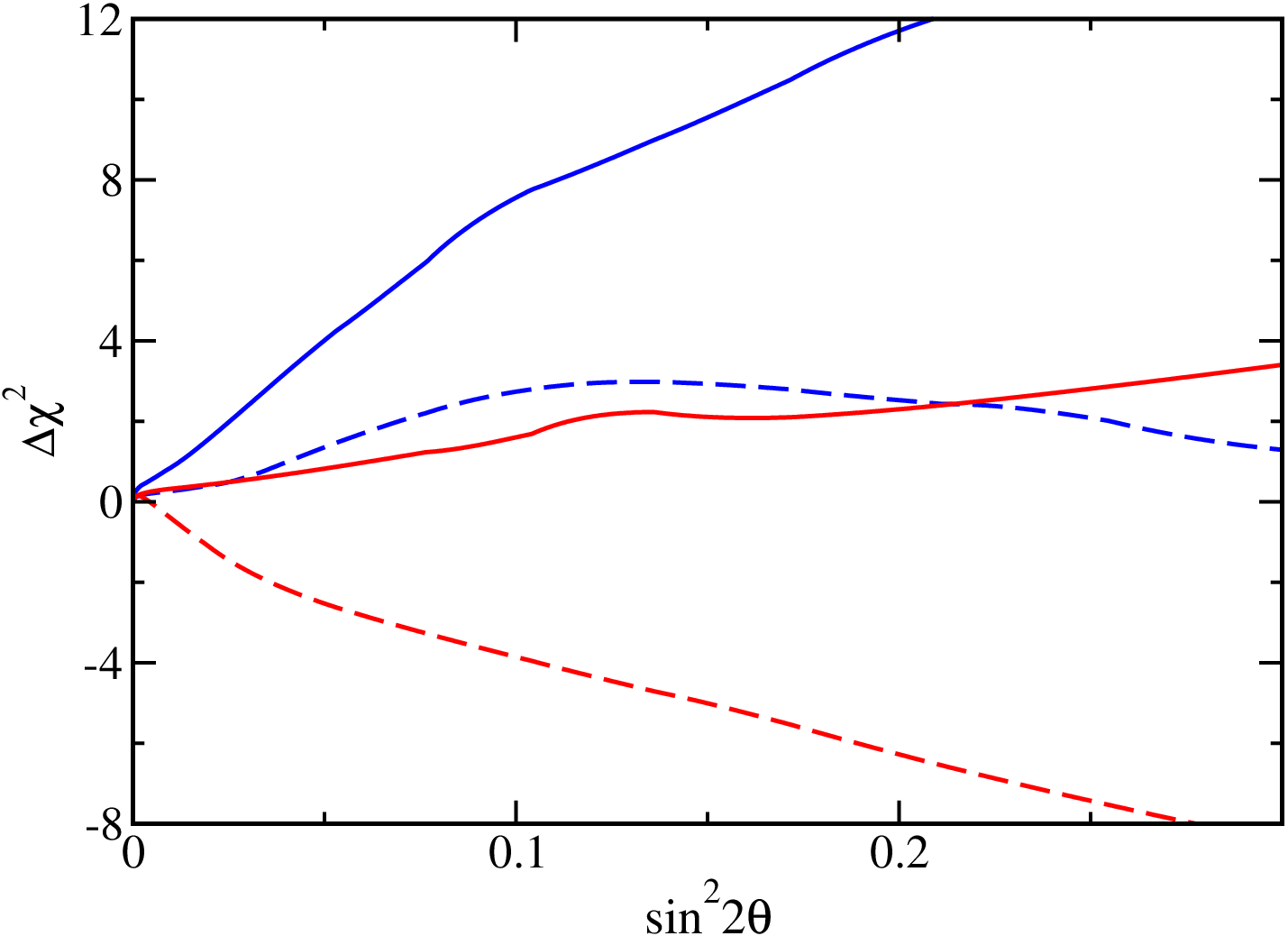}\hspace{0.1in}\includegraphics*[width=3in]{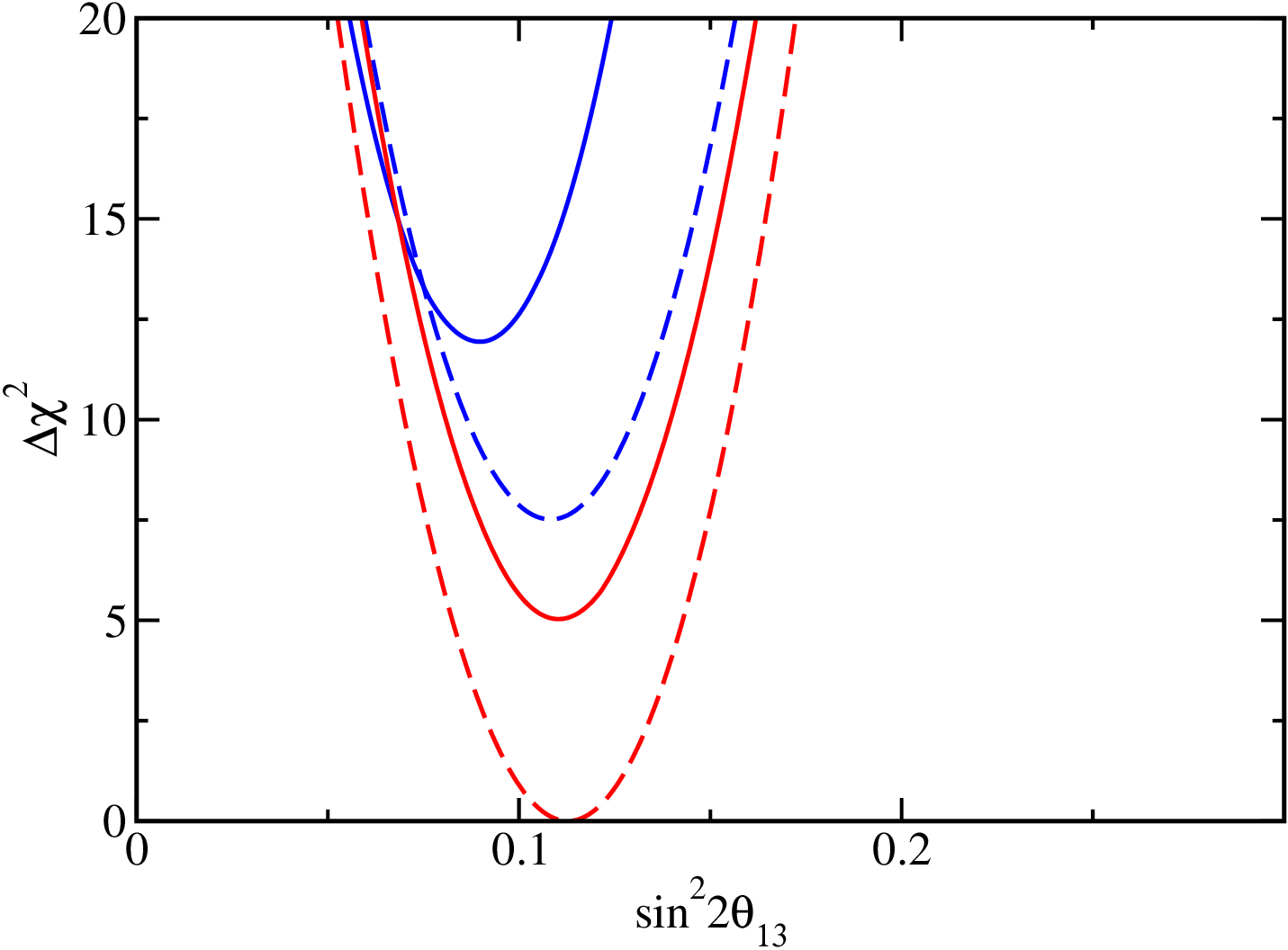}
\caption
{[Left] $\Delta\chi^2$ versus $\sin^22\,\theta_{13}$ for the Super-K atmospheric experiment.  [Right] $\Delta\chi^2$ versus $\sin^22\,\theta_{13}$ for all experiments. The curves are coded as in Fig.~\protect\ref{fig2}. }
\end{center}
\label{fig5}
\end{figure}

The implications of the Super-K atmospheric experiment \cite{17} for $\theta_{13}$ have been discussed in Ref.~\cite{5}. The features discussed there are visible in Fig.~\ref{fig5}. There are two subsets of the atmospheric data that play a role in determining $\theta_{13}$ and in discerning the hierarchy. One is data in the 3 to 10 GeV range with neutrinos passing through matter. The MSW interaction with matter produces resonances in this energy region for the normal hierarchy, but not the inverse hierarchy. The data do not indicate the existence of these resonances, and thus, for normal hierarchy, atmospheric data constrain the magnitude of $\theta_{13}$ with no such constraint for the inverse hierarchy. The second subset of the atmospheric data that constrains $\theta_{13}$ is data at large $L$ and small $E$ in the range of $L/E$ where there is interference between the oscillations arising from the two mass-squared differences.  These interference terms are linear in $\theta_{13}$. Combining the quadratic resonance terms and the linear terms, it was found that for normal hierarchy and positive $\theta_{13}$ the two terms are of the same sign and produced a constraint on $\theta_{13}$ that sets the bound more stringently than did the Chooz experiment. However, for normal hierarchy and negative $\theta_{13}$ the two terms are of opposite sign and of nearly equal magnitude and thus tend to cancel. For the inverse hierarchy one has only the linear in $\theta_{13}$ terms, and they favor the negative hierarchy. This behavior can be seen in the graphs depicting $\Delta\chi^2$ versus $\theta_{13}$ found in Ref.~\cite{5}. We note that these results are similar to those presented in Fig.~[24] of Ref.~\cite{18} realyzing that their $\cos\delta=-1$ branch is our negative $\theta_{13}$ branch. 
\vfill\eject
\section{Conclusions}
\begin{table}[t]
\caption{Results for a global analysis for the four combinations of hierarchy and sign of $\theta_{13}$. The last column gives the percent probability that the choice of hierarchy and $\theta_{13}$ are correct.}
\begin{center}

\begin{tabular}{lcccccc}
\br
hierarchy&sign $\theta_{13}$&$\sin^22\,\theta_{13}$&\% probable\\
\mr
normal&$+$&$0.090\pm0.020$&0.2\%\\
normal&$-$&$0.108\pm0.023$&2.2\%\\
inverse&$+$&$0.093\pm0.022$&7.1\%\\
inverse&$-$&$0.113\pm0.022$&90.5\%\\
\br
\label{tab3}
\end{tabular}
\end{center}
\end{table}

We present our results for an analysis that includes all the experiments in Fig.~\ref{fig5} where we show $\Delta\chi^2$ for the four cases of hierarchy and sign of $\theta_{13}$. The location of the minima are found to be only slightly affected by the choice of hierarchy or sign of $\theta_{13}$. 
For the now known value of $\sin^22\,\theta_{13}\approx 0.1$, atmospheric data distinguish between hierarchy and the sign of $\theta_{13}$ at a not negligible level. The normal hierarchy, positive $\theta_{13}$ case is disfavored, while the inverse hierarchy, negative $\theta_{13}$ case is favored with the other two cases lying in the middle. This is made quantitative in Table~\ref{tab3}. By marginalizing over $\sin^22\theta_{13}$ we calculated the relative probability of each of the four cases. The probability that normal hierarchy, positive $\theta_{13}$ is correct is 0.2\%, a fairly strong indication that it is not correct. The inverse hierarchy, negative $\theta_{13}$ case is 90.5\% probable that it is correct, a not negligible preference, while the normal hierarchy, negative $\theta_{13}$ and the inverse hierarchy, positive $\theta_{13}$ cases are 2.2\% and 7.1\% probable, respectively.

We compare our best fit value $\sin^22\theta_{13}=0.113$ for negative hierarchy, negative $\theta_{13}$ to $\sin^22\theta_{13}=0.103$ found for normal hierarchy, positive $\theta_{13}$ or negative hierarchy, negative $\theta_{13}$ from Ref.~\cite{10} and to $\sin^22\theta_{13}=0.094$ found for positive or negative hierarchy and negative $\theta_{13}$ for Ref.~\cite{19}. There is disagreement on the sign of $\theta_{13}$ and the hierarchy question. We find a larger hierarchy difference than do the other two. Since the present T2K and MINOS experiments have little preference for the hierarchy and the sign of $\theta_{13}$, the differences between the analyses arise from the atmospheric analysis. 

These differences may be resolved by future Super-K atmospheric data. Although the Super-K data is statistically significant on the whole, the role of $\theta_{13}$ in the analysis is a small effect. For example, the bins that indicate the presence or absence of the MSW resonance, are statistically not too significant. Further, T2K and NO$\nu$A have not produced data of statistical significance. With the large value for $\theta_{13}$ their future data has great promise to help resolve this situation.

\ack The work of D.J.E. is supported, in part, by US Department of Energy Grant No. DE-FG02-96ER40975, the work of B.K.C. is supported, in part, by US Department of Education Grant No. P200A090275, and the work of J.E-R. is supported, in part, by CONACyT, Mexico.
\vfill\eject

\end{document}